\documentclass[sigconf]{acmart}
\usepackage[ruled,vlined,linesnumbered]{algorithm2e}
\DeclareMathOperator*{\argmax}{argmax}
\newtheorem{theorem}{Theorem}
\newtheorem{lemma}[theorem]{Lemma}
\usepackage{nicematrix}
\usepackage{subcaption}

\usepackage{pgfplots}
\pgfplotsset{compat=newest}
\usepackage{tikz}

\newcommand{\plotfile}[1]{
    \pgfplotstableread{#1}{\table}
    \pgfplotstablegetcolsof{#1}
    \pgfmathtruncatemacro\numberofcols{\pgfplotsretval-1}
    \pgfplotsinvokeforeach{1,...,\numberofcols}{
        \pgfplotstablegetcolumnnamebyindex{##1}\of{\table}\to{\colname}
        \addplot table [y index=##1] {#1}; 
        \addlegendentryexpanded{\colname}
    }
}

\AtBeginDocument{%
  \providecommand\BibTeX{{%
    \normalfont B\kern-0.5em{\scshape i\kern-0.25em b}\kern-0.8em\TeX}}}

\begin{document}

\title{Doing away with Cameras: A Dynamic Obstacle Tracking Strategy for Proactive Handoffs in Millimeter-wave Networks}

\author{Rathindra Nath Dutta}
\authornote{All authors contributed equally to this manuscript}

\affiliation{%
  \institution{Indian Statistical Institute}
  \city{Kolkata}
  \country{India}
}

\author{Subhojit Sarkar}
\authornotemark[1]
\affiliation{%
  \institution{Indian Statistical Institute}
  \city{Kolkata}
  \country{India}
}

\author{Sasthi C. Ghosh}
\authornotemark[1]
\affiliation{%
  \institution{Indian Statistical Institute}
  \city{Kolkata}
  \country{India}
}

\begin{abstract}
Stringent line-of-sight demands necessitated by the fast attenuating nature of millimeter waves (mmWaves) through obstacles pose one of the central problems of next generation wireless networks. These mmWave links are easily disrupted due to obstacles, including vehicles and pedestrians, which cause degradation in link quality and even link failure. Dynamic obstacles are usually tracked by dedicated tracking hardware like RGB-D cameras, which usually have small ranges, and hence lead to prohibitively increased deployment costs to achieve complete coverage of the deployment area. In this manuscript, we propose an altogether different approach to track multiple dynamic obstacles in an mmWave network, solely based on short-term historical link failure information, without resorting to any dedicated tracking hardware. After proving that the said problem is NP-complete, we employ a greedy set-cover based approach to solve it. Using the obtained trajectories, we perform proactive handoffs for at-risk links. We compare our approach with an RGB-D camera-based approach and show that our approach provides better tracking and handoff performances when the camera coverage is low to moderate, which is often the case in real deployment scenarios.
\end{abstract}




\maketitle

\section{Introduction}
The ever increasing usage in internet-enabled devices promises to soon drain the scant traditional spectrum that is still available. The high bandwidth provided by millimeter wave (mmWave) communication, coupled with its vast unused spectrum (more than 200 times the currently allocated spectrum \cite{5783993}), makes it an attractive candidate to cater to such rising demands.
mmWaves have short wavelengths in the order of millimeters, and they consequently suffer from high attenuation through free space, and even more through obstacles. With the advent of massive MIMO, techniques like beamforming and large array gains can be used to overcome the high free space path loss of mmWaves; however, their penetration loss remains a major bottleneck. A static network topology would have eased the problem, with links being assigned by ensuring a line-of-sight (LOS) between communicating devices (either directly, or via relays/reflecting devices). Presence of dynamic obstacles like vehicles and even pedestrians, however, causes network topology to be time varying, which in turn makes the problem of mmWave link allocation far more challenging. Collogne et al. \cite{collonge2004influence} investigated the effects of human mobility on 60 GHz links, and concluded that the fading amplitude to be quite high.  It has been reported that human blockages cause up to a $20$-$40$ dB attenuation \cite{7881087, 6958971, 1374945}, while for a tinted car window the attenuation is $30$-$35$ dB \cite{9411094}. It has also been reported that such obstruction causes intermittent outages, decay in system throughput, and degrades overall user experience \cite{infocom}. Additionally, a number of unnecessary handovers are often introduced because of dynamic obstacles \cite{9109717}, which further add to system overhead. A comprehensive study on effects of pedestrian blockages, and subsequent mobility management, and network performance can be found in \cite{10.1145/3416012.3424621}.

The classical way to actively track dynamic obstacles is by deploying additional hardware like LiDARs \cite{LIDAR1} and cameras \cite{oguma2016proactive, charan2021vision,infocom}. After predicting future blockages, the approach is to handoff the communication to alternative base stations (BSs) as in \cite{infocom}. The primary drawback of deploying additional hardware like cameras, apart from the usual privacy concerns \cite{4658783}, is the considerable cost overhead that has to be borne by the service providers, and subsequently by the end users. Additionally, RGB-D cameras typically require small ranges for accurate measurements \cite{microsoft}, thereby requiring a very high number to be deployed, to achieve complete coverage. Furthermore, the training overhead of machine learning based approaches on the data procured from such tracking hardware are sometimes prohibitively high \cite{infocom}, which raises the questions about their scalability.

\begin{figure}[t]
    \centering
    \includegraphics[width=0.9\linewidth]{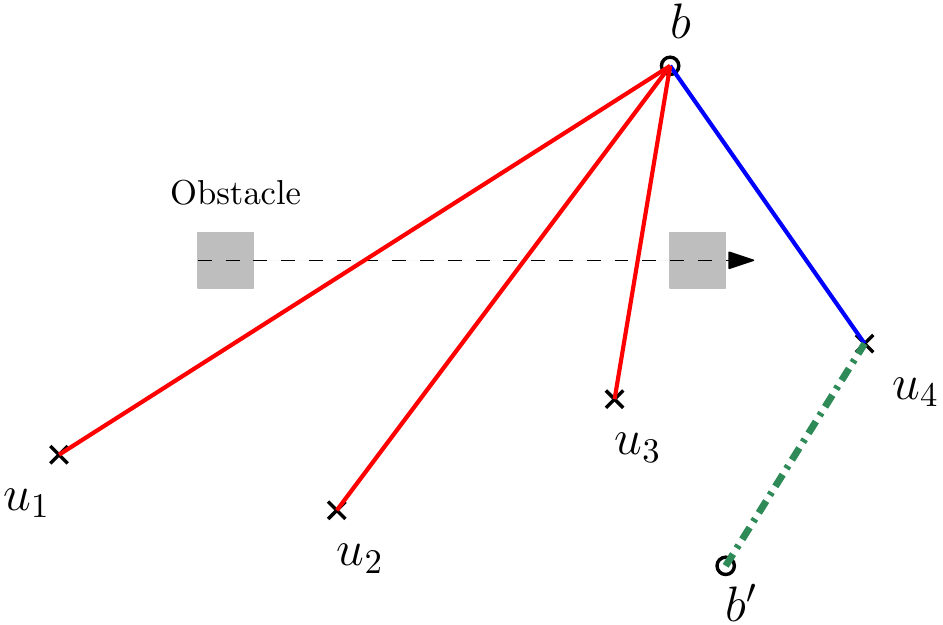}
    \caption{A dynamic obstacle blocking a few links}
    \label{fig:intro}
\end{figure}

In this manuscript, we approach the problem of tracking dynamic obstacles in an mmWave network without resorting to any dedicated tracking hardware. Given the usual network infrastructure, and short term historical link failure information, we attempt to obtain the trajectories of dynamic obstacles, and use them to achieve proactive handoffs with an aim to avoid obstacle induced link quality degradation. We demonstrate this target through a toy example in Fig. \ref{fig:intro}, where there are two BSs $b$ and $b'$ denoted by black circles, and $4$ user equipment (UE) which are denoted by black crosses. A single dynamic obstacle denoted by the grey square is moving along the dashed line, and has already broken the links $\{u_1-b\}$, $\{u_2-b\}$, and $\{u_3-b\}$ links (each denoted in red). There is, however, the $\{u_4-b\}$ link (denoted in blue) that is likely to be broken by the same dynamic obstacle. We attempt to preempt failure of this link by estimating the trajectory of the said obstacle, and thereafter performing a handoff to BS $b'$; thus, link $\{u_4-b'\}$ (shown in green) is formed.  All this is performed without using any dedicated tracking hardware. This is the core idea of the paper.
%
Formally, our main contribution in this paper can be summarised as follows:
\begin{itemize}
    \item We have embarked upon, what we believe, is the first attempt to track multiple dynamic obstacles that obstruct LOS transmission in a mmWave network, without resorting to any dedicated tracking hardware like RGB-D cameras.
    \item An integer linear program (ILP) is developed for the dynamic obstacle tracking problem (DOTP), and the problem is proved to be NP-complete.
    \item Using short term historical link failure information, we provide a greedy set cover based algorithm to obtain the trajectories of dynamic obstacles, and use them to achieve proactive handoffs before links are actually disrupted.
    \item We compare our proposed approach with an RGB-D camera based approach. We show that for low to moderate camera coverage, our approach produces better obstacle tracking, and subsequently manages to avoid more link failures. We emphasize that since ours is merely a predictive approach, tracking through complete camera coverage will definitely outperform our method. However, such ubiquitous tracking would no doubt be accompanied by excessively high expenses, which would possibly make it infeasible in practice.
\end{itemize}
It would be pertinent to point out here that the aim of this paper is primarily to demonstrate the viability of using link failure information to track dynamic obstacles, without the need for any tracking hardware. The approaches presented here are very basic, and we hope that this idea sparks interest in hardware independent obstacle tracking approach, which can possibly help in lowering infrastructure costs in next generation networks.

The rest of the paper is arranged as follows. We describe the state of the art in obstacle prediction in Sec. $2$, and present the considered system model along with its assumptions in Sec. $3$. We present a mathematical formulation of the problem at hand in Sec. $4$, and prove its hardness. We describe the proposed greedy algorithms and the simulations performed in Sections $5$ and $6$ respectively. Finally, concluding remarks are given in Sec. $7$.
\section{Background and Related Work}
Obstacle induced link failures are typically dealt in one of two ways. In the reactive approach as in \cite{5449837, 9275355}, the handoff takes place only after the blocking has taken place,
while in the proactive approach, link failures are predicted beforehand, and corrective measures taken accordingly. The proactive approach usually
involves deploying dedicated tracking hardware, like RGB-D cameras, radars, or LiDARs. Nishio et al. \cite{7157963} used an RGB-D camera to predict the location and mobility of human obstacles, and implemented a traffic mechanism that stops communication on the soon-to-be-blocked path, thereby freeing up that channel for use elsewhere. Authors in \cite{8792137} leveraged camera imagery and convolutional long short-term memory (LSTM) based machine learning to achieve proactive handoffs. More recently, Charan et al. \cite{charan2021vision} proposed a machine learning based framework that deployed RGB cameras at the BSs to track dynamic obstacles, and achieve proactive handoffs preventing link blockages. Oguma et al. \cite{oguma2016proactive} also uses RGB-D camera images to localise pedestrians and estimate their mobility to predict link blockages; subsequently, proactive handoffs are carried out before the human can block the link, thereby avoiding link failure. More recently, Zhang et al. \cite{9798197} demonstrated the link switching in an mmWave environment, by tracking obstacles using a stereo camera for well-lit environment, and a LiDAR for dark environment. A reinforcement learning based handover mechanism was proposed by Yoda et al. \cite{infocom}, which used a dedicated human tracking module.
The problem with hardware like RGB-D cameras is that they typically have ranges of a few meters ($0.5$ m to $3.5$ m) \cite{microsoft}; indeed, \cite{8108570} demonstrates an obstacle tracking approach using an RGB-D camera, where the obstacle distance is in the range of few meters. As such, their extensive deployment throughout the coverage area remains a big question. There have been works \cite{7564769} that assume complete knowledge of the mobility of obstacles in the coverage area, an assumption which infeasible in real life. Similarly, \cite{8792137} considers a model where the communication path is always within the field of view of an RGB-D camera, which might not always be the case. Moreover, learning based approaches often require prohibitively large training overhead (for example, \cite{infocom} uses $10^{10}$ tuples just for a $4$ m $\times$ $5$ m area, and two BSs), which might be unscalable.

A slightly different way of dealing with dynamic obstacle induced link failures is via multi-connectivity \cite{7414142, 7528494}, an extension of the dual connectivity \cite{7959177} that has already been proposed in LTE. Here, multiple BSs are associated with the same UE;
if one link fails due to an obstacle, communication continues via the others. Another way that uses multiple beams was proposed recently by Hersyandika et al. in \cite{hersyandika2022guard}. It used an extra guard beam to predict incoming obstructions, in order to protect the main communication beam. However, while multi-connectivity for all users has been recently reported to decrease network throughput \cite{10038560}, using a separate guide beam would decrease spectral efficiency and increase energy demands. Authors in \cite{10.1007/978-3-031-35299-7_6} had proposed a method that uses the list of BSs ``visible'' from each UE, to track a single dynamic obstacle in the coverage area, an approach which, apart from its stringent continuous beamforming requirements, fails to handle multiple dynamic obstacles. 

Link failures have been previously used by Sun et al. \cite{9109717} to track the location and trajectory of a UE. For a given UE, they essentially used a signal space partitioning scheme based on its LOS information with the BSs, and proposed a multi-armed bandit approach to achieve efficient handovers. Motivated by their idea, we attempt to use short term historical link failure data to track the dynamic obstacles. The challenge lies in the fact that while UEs can report their spatiotemporal details to a BS, the same is not true for dynamic obstacles as they move independently outside the purview of the BS. The problem becomes harder when considered without any dedicated tracking hardware to obtain such details. To the best of our knowledge, there exists no previous work that attempts to deal with the multiple dynamic obstacle tracking problem (DOTP) in a mmWave network, without resorting to additional tracking hardware. In this work, using the knowledge of short term historical link failures, we try to obtain the possible obstacle trajectories, and subsequently avoid link failures by preemptive handoffs.

%
\section{System Model}\label{sec:model}
We consider a service area having a central LTE BS, and a set $\mathcal{B}$ of small cell mmWave BSs distributed uniformly at random within the service area. All the BSs are connected to each other by means of a high speed backhaul network. Inside the service area, there are some UEs whose accurate locations are available at the LTE BS, and who wish to have a high speed link with any mmWave BS. We discretize the time into slots $0, \Delta, 2\Delta, \dots$, where each slot is of duration $\Delta$.

%
%
\paragraph{Path Loss Modelling: }
The path loss $PL(d_{i,j})$  between two devices $i$ and $j$ at a distance $d_{i,j}$ from each other is calculated as \cite{6834753}
\begin{equation}
PL(d_{i,j}) = \mathrm{A} + 10 \mathrm{B} log_{10}(d_{i,j}) + \zeta.
\end{equation}
Here, $\mathrm{A}$ and $\mathrm{B}$ are the pathloss parameters, and $\zeta$ is a log-normal random variable with zero mean, and variance $\sigma^2$.

Each UE can communicate with an mmWave BS if there is an LOS path between the two, and the free space path loss is lesser than a threshold. Though non line of sight (NLOS) mmWave communication  has been reported in some works like \cite{6398884}, the corresponding path loss exponents double as compared to LOS \cite{rappaport2014millimeter}; hence in this work, we assume communication can happen over LOS paths only. If an LOS does not exist between a UE and any one of the mmWave BSs, the LTE BS steps in to provide traditional sub 6 GHZ service. Each mmWave link can be represented by a line segment $\{u-b\}$, the end points being the positions of the UE $u$ and the mmWave BS $b$ respectively.
\paragraph{Obstacle Modelling: }
There are a number of dynamic obstacles moving independently inside the coverage area. For tractability, we consider these dynamic obstacles to be cuboids, whose 2-D projections are squares.
The obstacles move along a straight line for a time epoch $T$ (elucidated in the next paragraph). This is a reasonable assumption since in real life, a car, for example, changes directions rather infrequently, and $T$ is a small period in the order of few seconds. Due to the weakly penetrating nature of mmWaves, these obstacles cause significant degradation in mmWave link quality, causing link failure, which is reported to the corresponding mmWave BS.
Links operating in the mmWave spectrum can fail due to multiple causes, such as obstacles on transmission path, imperfect channel state information, or high interference from neighbouring links. Since primary focus of this work is to track the dynamic obstacles from past link failures, we assume that link failures occur due to the presence of dynamic obstacles only.
%
%
%
\begin{figure}
    \centering
    \includegraphics[scale=0.7]{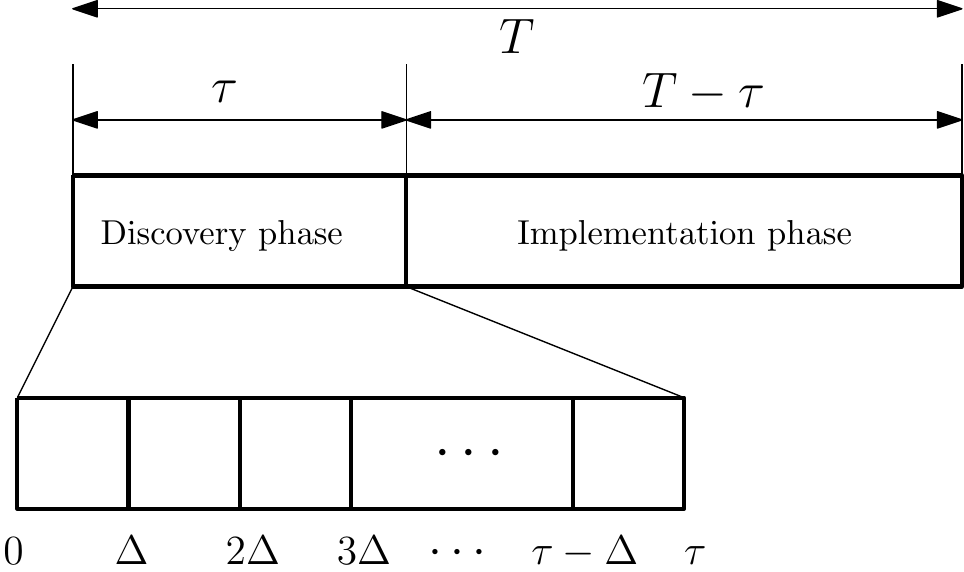}
    \caption{A time epoch $T$}
    \label{fig:tau}
\end{figure}
\paragraph{Discovery and Implementation Phases: }
We divide up the time into epochs of duration $T$. Each time epoch is further divided up into two phases, namely the \textit{discovery} phase of duration $\tau$, and the \textit{implementation} phase of remaining duration $T-\tau$, as shown in Fig. \ref{fig:tau}. Each mmWave BS keeps a record of the set of link failures that it witnessed in the discovery phase of an epoch. At the end of the discovery phase, each mmWave BS utilizes the corresponding link failure information to figure out the possible obstacle trajectories within its purview, which are then used to predict potential link failures in the implementation phase. After each time epoch $T$, the process starts afresh. As mentioned in the previous paragraph, the dynamic obstacles do not change their direction over an epoch. The tracking algorithm runs at each mmWave BS after the completion of the discovery phase.

\section{Problem Formulation}\label{sec:ILP}
%
Let $\mathcal{L}$ be the set of all blocked links associated with an mmWave BS after the discovery phase of a certain epoch. Let us assume that the upper bound on the number of dynamic obstacles is $K$.

In order to write the considered DOTP in a mathematical form, we introduce a set of binary indicator variables $X_k \in \{0,1\}$ for all $0\le k\le K$ defined as follows:
\begin{equation}
    X_k = \begin{cases}
        1 & \text{when obstacle $k$ causes at least one link failure}\\
        0 & \text{otherwise}
    \end{cases}\nonumber
\end{equation}
Now, our objective is to find the minimum number of obstacles accounting for all link failures. Thus, the objective function can be given by:
\begin{equation}
    minimize: \sum\limits_{k=1}^{K} X_k\label{eqn:obj}
\end{equation}
Now let us introduce another set of binary indicator variables $Y_{k,l} \in \{0,1\}$ for all $1\leq k\leq K$ and $1\leq l \leq |\mathcal{L}|$ defined as follows:
\begin{equation}
    Y_{k,l} = \begin{cases}
        1 & \text{when $l$-th link is failed due to $k$-th obstacle}\\
        0 & \text{otherwise}
    \end{cases}\nonumber
\end{equation}
Clearly we need to have,
\begin{equation}
    \sum\limits_{k=1}^{K} Y_{k,l} \ge 1 ~~\forall l\label{eqn:cover}
\end{equation}
At time instant $t$, a link $l$ is failed due to an obstacle $k$, only if the trajectory line of obstacle $k$ intersects the line segment representing the link. For notational brevity, from now onward we refer to the trajectory of obstacle $k$, and the line segment representing the link $l$, simply by the indices $k$ and $l$ respectively. 
A link $l$ is defined by its two end points $l_b$ and $l_e$. Now an obstacle $k$ intersects a link $l$ at point $P_{k,l}$ if and only if $k$ partitions the line segment $l$ in the ratio $\alpha_{k,l}$ and $1-\alpha_{k,l}$, where $0\le \alpha_{k,l}\le 1$ . The coordinate of $P_{k,l}$ can be computed from $l$ and $\alpha_{k,l}$ as follows.
\begin{gather}
P_{k,l}[x] = \alpha_{k,l} l_b[x] + (1-\alpha_{k,l}) l_e[x]\label{eqn:x_1}\\
P_{k,l}[y] = \alpha_{k,l} l_b[y] + (1-\alpha_{k,l}) l_e[y]\label{eqn:y_1}\\
P_{k,l}[t] = t \text{ such that $l\in \mathcal{L}_t$}
\end{gather}
Here for a point $P$, $(P[x], P[y])$ denotes its spatial coordinates on the Cartesian plane, and $P[t]$ denotes the time when this point on the link $l$ is being considered and $\mathcal{L}_t$ is the set of all such links which are blocked at time $t$. Thus $\mathcal{L} = \cup_{1\le t \le \tau} \mathcal{L}_t$. Now the intersection point $P_{k,l}$ must also lie on the trajectory line $k$, and it must therefore satisfy the equation of the line $k$ given by
\[
    \frac{x - x_k}{\delta x_{k}} = \frac{y - y_k}{\delta y_{k}} = \frac{t-0}{\delta t_{k}}
\]
Here $(x_k,y_k,0)$ is the initial point on the line $k$ at time $t=0$. The parameters $\delta x_{k}, \delta y_{k}$ and $\delta t_{k}$ are the intercept values with $x, y$ and $t$ axes respectively. Thus, we have $x = x_k + t\frac{\delta x_{k}}{\delta t_{k}}$ and $y = y_k + t\frac{\delta y_{k}}{\delta t_{k}}$. Since for a line, $\frac{\delta x_{k}}{\delta t_{k}}$ and $\frac{\delta y_{k}}{\delta t_{k}}$ are constants, they can be dealt with only two variables, namely $A_k$ and $B_k$ respectively, in our mathematical program. Thus, whether the point $P_{k,l}$ lies on the line $k$ or not, can be encoded by the following linear constraints $\forall k,l$:
\begin{gather}
    P_{k,l}[x] = x_k + P_{k,l}[t]~A_k\label{eqn:x_2}\\
    P_{k,l}[y] = y_k + P_{k,l}[t]~B_k\label{eqn:y_2}
\end{gather}
Note that here, $P_{k,l}[t] = t$, is a constant specified by the link $l\in\mathcal{L}_t$. Now combining \eqref{eqn:x_1} with \eqref{eqn:x_2}, and \eqref{eqn:y_1} with \eqref{eqn:y_2} we have the following two linear constraints:
\begin{gather}
    \alpha_{k,l} l_b[x] + (1-\alpha_{k,l}) l_e[x] = x_k + P_{k,l}[t]~A_k\label{eqn:x_3}\\
    \alpha_{k,l} l_b[y] + (1-\alpha_{k,l}) l_e[y] = y_k + P_{k,l}[t]~B_k\label{eqn:y_3}
\end{gather}
Note here $\alpha_{k,l}, x_k, y_k, A_k, B_k$ all are optimization variables and the rest are constants. We must also ensure $0\le \alpha_{k,l}\le 1$ whenever $Y_{k,l} = 1$. This can be encoded as a linear constraint by introducing a large positive constant $M$, denoting positive infinity as follows:
\begin{equation}
    -(1-Y_{k,l})M \le \alpha_{k,l} \le 1 + (1-Y_{k,l})M\label{eqn:alpha}
\end{equation}
The integrality constraints are given by
\begin{equation}
    X_k, Y_{k,l} \in \{0,1\}\label{eqn:binary}
\end{equation}
Thus the ILP is given by the objective function \eqref{eqn:obj} and constraints \eqref{eqn:cover}, \eqref{eqn:x_3}, \eqref{eqn:y_3}, \eqref{eqn:alpha} and \eqref{eqn:binary}.

We next prove that the problem of detecting multiple obstacle trajectories is NP-complete.
\begin{lemma}
DOTP is NP-complete.
\end{lemma}
\begin{proof}
To show that DOTP is NP-complete, we choose the point-line-cover (PLC) \cite{PLC2} problem as the candidate for reduction. In PLC, given $n$ points and an integer $K$, we need to decide whether there exists $K$ straight lines such that all points are covered. Here, a line $L$ is said to cover a point $p$ if and only if $p$ lies on $L$. PLC is a known NP-complete  problem \cite{PLC2}.
Given an instance $I=(P,K)$, where $P=\{p_1, p_2, \cdots, p_n\}$, of PLC, we apply the following reduction. For every point $p_i\in P$, we create two points $p_i'$ and $p_i''$, both having coordinates same as that of $p_i$. Moreover, $(p_i',p_i'')$ forms a link of zero length at time $t_i$ for our DOTP. Now suppose, there exists a deterministic polynomial time algorithm $\mathcal{D}$ for DOTP. Then for a given instance $I'=(\mathcal{L},K)$ of DOTP, where $\mathcal{L}_{t_i}=\{(p_i',p_i'')\mid p_i\in P\}$, and $\mathcal{L}=\bigcup\limits_{i=1}^{n} \mathcal{L}_t$, $\mathcal{D}$ decides in polynomial time whether there exist $K$ lines such that all links are intersected by at least one of these lines. Now by construction, if a line intersects a link $(p_i',p_i'')$, it must also cover the original point $p_i$ in PLC, and vice versa. This means we have essentially solved instance $I$ of PLC in polynomial time. This is a contradiction. Therefore, DOTP is NP-hard, and such an algorithm $\mathcal{D}$ cannot exist unless P=NP.

Now to show DOTP is also in NP, consider the fact that given $K$ obstacle trajectories (lines), one can easily verify whether all links are blocked (intersected) by at least one of them. Thus, DOTP is NP-complete. 
\end{proof}

\section{Discovery and Implementation Algorithms}\label{sec:algo}
As mentioned in Section \ref{sec:model}, our approach works in two phases. While in the discovery phase we try to obtain the possible trajectories of the dynamic obstacles, in the implementation phase we apply this knowledge to achieve proactive handoffs with an aim to avert link failures.
We now present a greedy algorithm to obtain the set of obstacle trajectories $\mathcal{O}$, and follow it up with a simple handoff scheme to avert link failures.

\subsection{Dynamic Obstacle Tracking Algorithm}
We can model DOTP as a set cover problem as follows. Suppose we are given some candidate trajectory lines, each covering a subset of the universe $\mathcal{L}$. A trajectory line $k$ covers a link $l\in\mathcal{L}$ if $k$ intersects $l$. Then an optimal (minimum) set cover of this universe, essentially gives us the required solution of the DOTP. Recall that the set cover problem is a well known APX-hard problem \cite{apxset}. A greedy solution to set cover can be obtained by repeatedly selecting the set covering maximum number of yet-to-be-covered elements. This approach has an approximation ratio of $\log n$, where $n$ is the number of candidate sets. Thus given a set of possible trajectory lines, we already have an approximation algorithm that returns a minimal subset of trajectory lines covering all links.
The main challenge is to get this set of candidate trajectory lines, as there can be infinitely many possible lines intersecting just two links! However, in the following lemma, we show that we can always generate a finite set of such candidate lines.

\begin{lemma}
    If a line $L$ intersects the set of $m$ links $\mathcal{L} = \bigl\{\{u_1-b\}, \{u_2-b\}, \cdots, \{u_m-b\}\bigr\}$ associated with a BS $b$, there exists at least one $(u_i, u_j)$ pair ($1\leq i,j \leq m, i\neq j$) such that the line passing through $u_i$ and $u_j$ must also intersects all of these links.    
\end{lemma}

\begin{proof}
    Let us assume that an oracle has provided us with the actual trajectory line $L$ of the obstacle which causes the $m$ links to fail. We can always translate the given trajectory line parallel to itself till it reaches the location of any one UE, say $i$; let the new line parallel to $L$ and passing through $i$ be called $L'$. Thereafter, keeping $i$ as a pivot, we can rotate $L'$ till it touches another UE, say $j$. Let this line obtained after rotating $L'$ about $i$ be called $L''$. Notice that $L''$ is an extension of the line segment formed by joining $i$ and $j$. Furthermore, the translation and rotation operations have been done ensuring that all links are touched by the line $L''$, i.e., none of the links become uncovered. Hence $L''$ is a candidate trajectory.
\end{proof}
\begin{figure}
    \centering
    \includegraphics[scale=0.65]{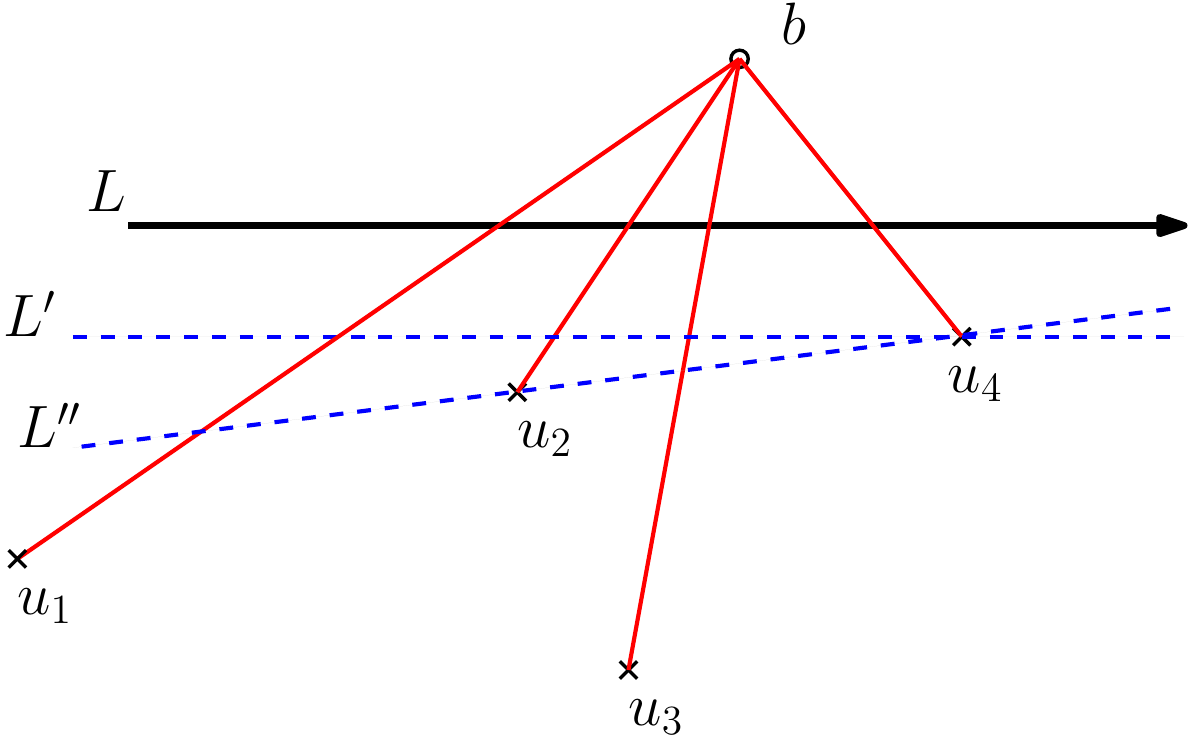}
    \caption{$L''$ is a candidate trajectory line that intersects all the blocked links}
    \label{fig:lemma}
\end{figure}
In Fig. \ref{fig:lemma}, $L$ represents the actual trajectory line. $L'$ is the translated line that passes through $u_4$ and parallel to $L$. After rotating $L'$, about $u_4$, we get $L''$ which passes through both $u_2$ and $u_4$, and intersects all the links in $\mathcal{L} = \bigl\{\{u_1-b\}$, $\{u_2-b\}$, $\{u_3-b\}$, and $\{u_4-b\}\bigr\}$.


Lemma $2$ actually gives us a way to obtain a finite candidate trajectory set. If we consider the set $\mathcal{C}$ of all lines passing through the all possible pairs of UEs associated with a BS, the required trajectory lines must be a subset of $\mathcal{C}$. As there can be $\binom{|\mathcal{L}|}{2}$ possible UE pairs, $|\mathcal{C}| = O(|\mathcal{L}|^2)$.
Moreover, the optimal solution must be the minimum subset of $\mathcal{C}$ covering entire $\mathcal{L}$.
Given a candidate line $c \in \mathcal{C}$ jointing two end points of two links from $\mathcal{L}$, we scan the entire $\mathcal{L}$, to find how many links are intersected by this line $k$, which gives us the set $\mathcal{C}_k \subseteq \mathcal{L}$, the set of links covered by this line. This can be computed in linear time with respect to $\mathcal{L}$. Once we obtain all such $\mathcal{C}_k$, we follow the footsteps of greedy set cover solution described above. This will return us at most $OPT \times \log (|\mathcal{L}|)$ many trajectories where $OPT$ is the size of optimal set cover solution.
This process is formalized into Algorithm \ref{algo:1}, which uses the \texttt{intersection()} procedure as described below.

\begin{algorithm}
Set $\mathcal{U}\leftarrow \mathcal{L}$\tcp{\small universe to cover}
Set $\mathcal{C}\leftarrow \emptyset$ \tcp{\small candidate trajectories}
Set $\mathcal{O}\leftarrow \emptyset$ \tcp{\small output trajectories which covers $\mathcal{U}$}
\tcc{\small generating candidate lines}
\For{$\{u-b\} \in \mathcal{L}$}{
    \For{$\{u'-b\} \in \mathcal{L}\text{ and }u'\ne u$}{
        $c \leftarrow (u, u')$ \tcp{\small candidate line}
        $\mathcal{C} \leftarrow \mathcal{C}\cup \{c\}$
    }
}
\tcc{\small apply greedy set cover}
\While{$\mathcal{U}\neq \emptyset$}{
    \For{$c\in \mathcal{C}$}{
        $I_c\leftarrow \{u \in \mathcal{U}\mid \texttt{intersection}(c, u)=1\}$\\
    }
    $k\leftarrow \argmax\limits_{c' \in \mathcal{C}}\{I_{c'}\}$\\
    $\mathcal{U} \leftarrow \mathcal{U} \setminus I_k$\\
    $\mathcal{O}\leftarrow\mathcal{O}\cup\{k\}$    
}
Return $\mathcal{O}$
\caption{Set Cover Based DOTP}\label{algo:1}
\end{algorithm}
\subsection*{\textup {{The \texttt{intersection($k$, $l$)} Procedure}:}}
Now, since link $l$ is defined by its two end points $l_b$ and $l_e$, a line $k$ intersects $l$ if and only if $l_b$ and $l_e$ lie on two opposite sides of $k$. Let the line equation of $k$ be written as $f_k(x,y) = 0$. Now, if we plug the coordinates of $l_b$ and $l_e$ into this equation of $k$, we would get values of opposite signs if and only if they lie on different sides of line $k$. This provides us with an efficient way to test for intersection: $k$ intersects $l$ if and only if $f_k(l_b)\times f_k(l_e) \le 0$. The subroutine returns $1$ if there is an intersection, $0$ otherwise.

Algorithm \ref{algo:1} takes input the set $\mathcal{L}$ after the time $\tau$, and outputs the set $\mathcal{O}$ of trajectories that intersects all links in $\mathcal{L}$. Since $|\mathcal{C}| = O(|\mathcal{L}|^2)$, and \texttt{intersection}() takes $O(|\mathcal{L}|^2)$, the overall running time of Algorithm \ref{algo:1} is $O(|\mathcal{L}|^4)$. 

\subsection{Proactive Handoff Algorithm}
Armed with the set $\mathcal{O}$ of reported trajectory lines, and the currently active set  of links $\mathcal{A}$, we can now proceed towards triggering handoffs for obstacle prone links in the implementation phase.
The algorithm begins off by taking input $\mathcal{O}$ and  $\mathcal{A}$, and outputs a set of new links $\mathcal{L}'$. Here, an active link $a\in \mathcal{A}$ is represented by a pair $\{u-b\}$. For each such active link $a$, we call the \texttt{intersection}() subroutine, and check for any possible intersection with trajectories in $\mathcal{O}$; if such a possibility exists, we trigger a handoff. For the UE $u$, we check whether an mmWave BS $b'\neq b$ exists, such that $\{u-b'\}$ is not intersected by any of the obstacle trajectories in $\mathcal{O}$. If such an mmWave BS $b'$ exists, we update the set $\mathcal{L}'$ with the new link $\{u-b'\}$. If no such mmWave BS exists, the LTE BS steps in to provide sub-6 GHZ service, and we update the set $\mathcal{L}'$ with the link $\{u-\texttt{LTE BS}\}$. This process is formalized into Algorithm \ref{algo:2s}. 
\begin{algorithm}
    Set $\mathcal{L}'\leftarrow \emptyset$ \tcp{\small new links}
    \For{$a=\{u-b\} \in \mathcal{A}$}{
        \For{$o \in \mathcal{O}$}{
            \If(\tcp*[h]{\small \hspace{-2pt}trigger handoff}){$\texttt{intersection}(a, o)=1$}{
                Set $flag\leftarrow 0$\\
                \For{$b'\in\mathcal{B}\mid b'\neq b$}{
                    \For{$o' \in \mathcal{O}$}{
                        \If{$\texttt{intersection}(\{u-b'\}, o')=0$}{
                            $\mathcal{L}'\leftarrow \mathcal{L}' \cup \{u-b'\}$\\
                            Set $flag\leftarrow 1$\\
                            break;
                    }
                }
                }
                \If{$flag=0$}{
                    $\mathcal{L}'\leftarrow \mathcal{L}' \cup \{u-\texttt{LTE BS}\}$
                }
                }
            }
        $\mathcal{A} \leftarrow \mathcal{A} \setminus a$

        }    
    
    Return $\mathcal{L}'$
    \caption{Proactive handoff algorithm}
    \label{algo:2s}
\end{algorithm}



\section{Simulation experiments}
To demonstrate the handoff performance of our algorithm and compare it with an RGB-D camera based method \cite{RGB-D}, we first select a smaller simulation setup as follows. We consider a service area of size $100$ m $\times$ $100$ m. There are $2$ mmWave BSs which can provide coverage to the UEs having LOS with any of them, and an LTE BS which provides ubiquitous coverage. There are a few obstacles of size $1$ m $\times$ $1$ m which are moving independently inside with a velocity chosen uniformly at random from [$0$, $10$] m/s. The obstacle can change its direction after an epoch $T$ of duration $5$ second. There are several UEs communicating with the mmWave BSs. The obstacles cause link failures, and the same is reported at the corresponding mmWave BS over the discovery phase. After the end of time $\tau=3$ second, each mmWave BS runs Algorithm
\ref{algo:1}, and determines a set of trajectories $\mathcal{O}$. Thereafter, Algorithm \ref{algo:2s} is run in a centralised manner at the LTE BS, and the set of links that are at risk of breaking due to the obstacles is determined. Handoff is triggered at such UEs in an attempt to provide uninterrupted service.
As is obvious, scarce UE density will lead to low number of link failures, leading to insufficient information, and inaccurate predictions. We demonstrate the effect of UE density in two confusion matrices; Table \ref{table:matrix1} shows the effect for $10$ failed links, while Table \ref{table:matrix2} shows the effect for $100$ failed links. It is evident that the algorithm performs poorly when the UE density is low, giving correct handoff requirements only $50\%$ of the time. However, the performance improves to above $80\%$ when the number of failed links is $100$. We point out here though, that achieving complete accuracy appears unlikely without dedicated tracking hardware. Indeed, the same obstacle may obstruct links associated with multiple mmWave BSs, which our algorithm would report as multiple trajectories, thereby introducing inherent false positives. In other words, in the absence of sufficient link failure information, the algorithm performs poorly.

\begin{table}
    \begin{subtable}[t]{0.48\linewidth}\centering
        {\small\begin{NiceTabular}{c@{\enskip}c@{\enskip}wc{1.2cm}wc{1.2cm}} 
                        &               & \Block{1-2}{\large Actual Obstruction}   \\[1ex]   
                        &               & Positive & Negative \\[1ex]
        \Block{2-1}{\rotate \large Predicted Obstruction}%
        \rule[-0.6cm]{0pt}{1.5cm}
                        & {\rotatebox[origin=c]{90}{Positive}}  & \Block[hvlines]{2-2}{}{\large 19.7}  & {\large 29.8} \\ 
        \rule[-0.6cm]{0pt}{1.5cm}
                        & {\rotatebox[origin=c]{90}{Negative}}  & {\large 20.8} & {\large 29.7} \\         
        \end{NiceTabular}}
        \caption{For 10 links}
        \label{table:matrix1}
    \end{subtable}\hfill
    \begin{subtable}[t]{0.48\linewidth}\centering
        {\small\begin{NiceTabular}{c@{\enskip}c@{\enskip}wc{1.2cm}wc{1.2cm}} 
                        &               & \Block{1-2}{\large Actual Obstruction}   \\[1ex]   
                        &               & Positive & Negative \\[1ex]
        \Block{2-1}{\rotate \large Predicted Obstruction}%
        \rule[-0.6cm]{0pt}{1.5cm}
                        & {\rotatebox[origin=c]{90}{Positive}}  & \Block[hvlines]{2-2}{}{\large 40.2}  & {\large 7.3} \\ 
        \rule[-0.6cm]{0pt}{1.5cm}
                        & {\rotatebox[origin=c]{90}{Negative}}  & {\large 7.5} & {\large 42} \\         
        \end{NiceTabular}}
        \caption{For 100 links}
        \label{table:matrix2}
    \end{subtable}
    \caption{Confusion Matrices}\label{table:both}
\end{table}

Next, we compare the tracking performance of our approach with an RGB-D camera based approach \cite{RGB-D}. As in \cite{RGB-D, 7157963}, we consider Microsoft Kinect cameras \cite{microsoft} to be the tracking devices. These cameras typically have a field of view of $90^\circ$, and range around few meters ($0.5$ to $3.5$ m) \cite{microsoft}. We vary the camera count, and deploy them uniformly at random inside the coverage area. We assume that obstacles whose trajectories lie fully within the coverage area of the cameras, are successfully tracked. As shown in Fig. \ref{fig:cameracover}, our proposed method (which is independent of the camera count) provides better tracking than the camera based approach, till about $100$ cameras are deployed for the considered $100$ m $\times$ $100$ m service area. Since ours is simply a prediction algorithm based on historical link failure data alone, sufficient coverage obtained by means of deploying a large number of RGB-D cameras will certainly outperform our approach.
In real life,  the inherent physical deployment limitations, along with existence of static obstacles, will push the number of cameras required to achieve full coverage to infeasible numbers.

\begin{figure}[ht]
    \centering
    \begin{tikzpicture}
        \begin{axis}[xlabel = No. of RGB-D cameras,
                    ylabel = Tracking capability,
                    width=0.85\linewidth,
                    height=5.5cm,
                    legend pos=south east]
            \plotfile{camera.dat}
        \end{axis}
    \end{tikzpicture}
    \caption{Obstacle tracking capability vs RGB-D camera count}
    \label{fig:cameracover}
\end{figure}
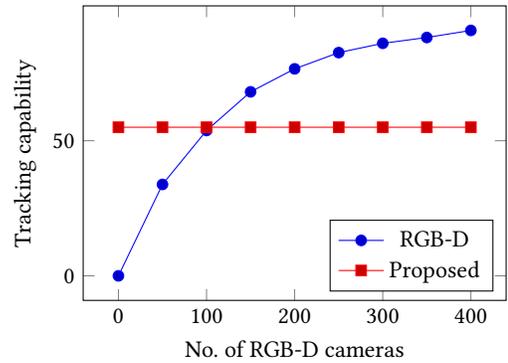

\begin{figure}[ht]
    \centering
    \begin{tikzpicture}
        \begin{axis}[xlabel = No. of RGB-D cameras,
                    ylabel = Handoff performance,
                    width=0.85\linewidth,
                    height=5.5cm,
                    legend pos=south east]
            \plotfile{handoff.dat}
        \end{axis}
    \end{tikzpicture}
    \caption{Handoff performance vs RGB-D camera count}
    \label{fig:handoffcameracount}
\end{figure}
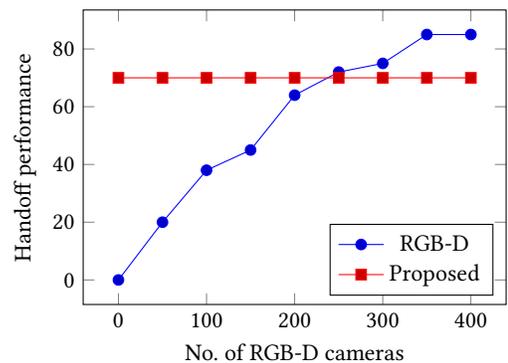

\begin{figure*}[th]
    \centering
    \includegraphics[page=4,width=\textwidth]{data/plot.pdf}\vspace{-0.5cm}
    \begin{subfigure}[b]{0.329\textwidth}
        \centering
        \includegraphics[page=1,width=\textwidth]{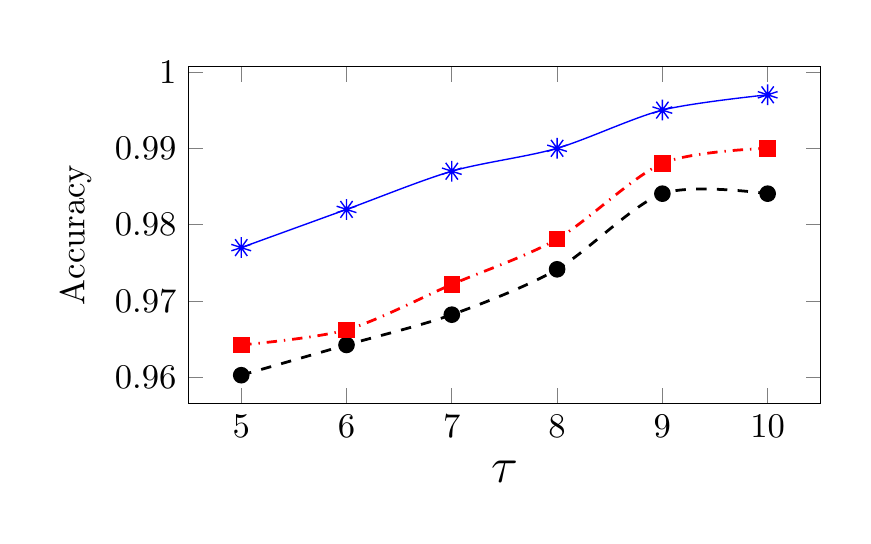}
        \vspace{-1cm}
        \caption{Accuracy}
    \end{subfigure}
    \begin{subfigure}[b]{0.329\textwidth}
        \centering
        \includegraphics[page=2,width=\textwidth]{data/plot.pdf}
        \vspace{-1cm}
        \caption{Sensitivity}
    \end{subfigure}
    \begin{subfigure}[b]{0.329\textwidth}
        \centering
        \includegraphics[page=3,width=\linewidth]{data/plot.pdf}
        \vspace{-1cm}
        \caption{Precision}
    \end{subfigure}
    \caption{Accuracy, Sensitivity and Precision vs. Discovery time $\tau$ for varying number of dynamic obstacles and link requests}
    \label{fig:sim1}
\end{figure*}

The comparison of handoff performance between the two methods mirrors the obstacle tracking performance as is evident in Fig. \ref{fig:handoffcameracount}. We define handoff performance as the percentage of links actually required handoffs, and were handed over to unobstructed mmWave BSs. For low camera count, the performance is worse since handoff requires tracking information at two stages; one to determine the at risk links, and the other to predict whether the newly allocated link will be obstructed. If the number of deployed RGB-D cameras is below $250$ for the considered coverage area of size $100$ m $\times$ $100$ m, our proposed approach clearly outperforms the camera based approach. If the camera density is huge (more than $250$ in this case), then the camera based approach performs better than ours.

To validate the performance of our proposed tracking approach in a real life scenario, we run simulations using the CRAWDAD taxi dataset \cite{taxi}, which contains GPS taxi traces in San Francisco Bay Area, USA.
We randomly select $15$ taxis from this dataset as our dynamic obstacles, and run our experiments considering a service area of size $5000$ m $\times$ $5000$ m. We consider a set of link requests generated uniformly at random over the service area, for the considered time period epoch $T=12$ second. We execute Algorithm \ref{algo:1} on the blocked links generated upto time $\tau$.

We measure the performance of our proposed scheme based on three metrics, namely \textit{accuracy}, \textit{precision}, and \textit{sensitivity}, which are defined as follows. 
\begin{itemize}
    \item \textit{accuracy}: the ratio of number of links correctly predicted according to their blocking status, to the total number of links under consideration.
    \item \textit{sensitivity}: ratio of the number of links correctly predicted to be ``blocked'', to the total number of actually blocked links. 
    \item \textit{precision}: ratio of the number of links correctly predicted as ``blocked'', to the total number of links predicted as ``blocked'' (both correctly, as well as incorrectly). 
\end{itemize}
We vary the duration of discovery phase $\tau$ from $5$ to $10$ seconds, and plot these three metrics in Fig. \ref{fig:sim1} for different number of obstacles and broken links. As is intuitive, all three metrics see an improvement with increase in discovery phase time $\tau$. However, if we increase $\tau$ indefinitely, the duration of the implementation phase shortens, thereby reducing the time for which the output of our tracking algorithm can be used for achieving uninterrupted service.


\section{Conclusion}
In this paper, we have taken a rather ambitious attempt at tracking multiple dynamic obstacles in an mmWave enabled coverage area, without deploying any sort of dedicated tracking hardware. The aim of this preliminary work was to demonstrate the viability of such an approach, and achieve subsequent proactive handoffs in an attempt to lower link failures. We proved the hardness of the problem, and modelled it as a version of the classical set cover problem, which we solved using the usual greedy approximation approach. Given the range constraints of RGB-D cameras, we show that unless near complete coverage is provided with huge number of such hardware, our approach performs better obstacle tracking, and subsequent handoffs. We further show that the performance of our predictive approach improves with high UE density (more link failure information), hypothesising that in sparse networks, we might need augmentation with some additional hardware to obtain satisfactory prediction, which can lead to new optimization problems. Since our tracking algorithm runs independently at each mmWave BS, in the implementation phase we can end up having more reported trajectories than the number of actual obstacles. This is because, the same obstacle which obstructed links associated with multiple mmWave BSs may be reported as multiple obstacles in our approach. One seemingly interesting way to improve the approaches in this work is by considering dedicated roads and sidewalks along which the dynamic obstacles are constrained to move upon. We have not assumed any such paths; it seems logical that such constraints will in fact, improve prediction performance. 
Such avenues will be explored in our future work.


\bibliographystyle{ACM-Reference-Format}
\bibliography{references}
\end{document}